# On the possibility of positive-ion detection in gaseous TPCs and its potential use for neutrinoless double beta decay searches in $^{136}$Xe

**Lior Arazi**

Physics Faculty, Weizmann Institute of Science, 234 Herzl St. Rehovot 7610001, Israel

E-mail: lior.arazi@weizmann.ac.il

**Abstract.** The neutralization of slow positive ions on solid surfaces can lead to the emission of secondary electrons in Auger-type processes. We discuss the possibility of harnessing such mechanisms to the detection of positive ions in gaseous TPCs. Applied to high pressure xenon, the proposed idea may enable reconstructing with high accuracy the topology of candidate neutrinoless double beta decay events of $^{136}$Xe without sacrificing the energy resolution of pure Xe gas. Candidate secondary electron emitters are discussed, as well as the expected efficiencies, challenges and potential pitfalls.

## 1. Introduction

Time projection chambers (TPCs) are a central tool in experimental high energy physics, nuclear physics and astroparticle physics, providing detailed three dimensional images of ionization tracks produced by charged particles passing through their sensitive volume. In the basic TPC scheme, ionization electrons liberated along the track are drifted towards a position-sensitive readout plane, where the track image is reconstructed. While in most applications this principle is sufficient, for some experiments traverse and longitudinal diffusion of the drifting electrons presents a serious limitation. This is especially true in cases where on the one hand the physics case requires resolving spatial features on the mm or sub-mm scale, and on the other hand the TPC dimensions must be large (meter-scale) to contain the full track or to allow for sufficient sensitivity, as in rare-event searches.

One way to suppress electron diffusion in gaseous TPCs is to apply a strong magnetic field parallel to the drift direction [1], but this is not always a practical or economic solution. Alternatively, one may use gas mixtures with small electron diffusion coefficients, but with a potential compromise of other properties of the detector, such as energy resolution [2]. A different approach, which may be applied if the event rate is sufficiently low, is to rely on *ions* to store the topology of the original track and transport it slowly and with minimal diffusion to an ion-sensitive readout plane at the endcap of the TPC. The Negative Ion TPC (NITPC) concept, developed by Martoff *et al* [3], implements this idea by using electronegative molecules to rapidly capture the electrons generated along the ionization track. The resulting negative ions are then drifted towards a readout plane where they are stripped of their extra electrons, which subsequently undergo avalanche multiplication in an appropriate amplification structure. The concept of a low-pressure NITPC is pioneered by the DRIFT program for directional dark matter detection [4], where few-mm long ionization tracks of low-energy (≲100 keV) nuclear recoils must be reconstructed with sub-mm accuracy to establish head-tail asymmetry. The collaboration's current detector, DRIFT-IId, uses a gas mixture of 30:10:1 Torr $CS_2$:$CF_4$:$O_2$, where fluorine atoms act as the target for WIMP scattering and $CS_2$ is the electronegative gas; avalanche multiplication is done on wires. An NITPC scheme is also under development for x-ray polarimetry of astronomical sources [5], using nitromethane ($CH_3NO_2$) as the electronegative gas and Gas Electron Multiplier (GEM) [6] structures for charge amplification.

The possibility of applying the NITPC concept to ton-scale searches for neutrinoless double beta decay ($0\nu\beta\beta$) in $^{136}$Xe was suggested by D. Nygren in [7]. The TPC, in this case, would consist of high pressure (~10-20 bar) Xe gas enriched in $^{136}$Xe with a small admixture of electronegative gas. The idea is to capture the track electrons by the electronegative molecules, drift the negative ions to the anode plane and there – since they drift very slowly - count them one by one, providing near-intrinsic energy resolution and accurate spatial information. A main challenge in realizing this concept is the identification of a suitable electronegative molecule, with electron affinity high enough to keep the extra electron bound during the drift, and low enough to release it in the amplification region at the anode plane. A second possible complication is the loss of the primary scintillation (and hence $t_0$) signal due to UV absorption either by the electronegative admixture itself or by other additive molecules which may be required for gain stabilization. As suggested in [7], $t_0$ may possibly be recovered using 'echo' signals generated by positive Xe (and Ba) ions on the cathode through the release of secondary electrons, which would be quickly captured and drifted to the anode by the electronegative molecules.

In this contribution we look (for now – theoretically) into the possibility of *positive* ion detection in TPCs as a complementary approach to suppress electron diffusion. The idea of extracting spatial information about the track from its positive ions is not new, but has so far been applied only in very low-pressure scenarios. Positive ion counting, relying on differential pumping of low pressure gas (< 1 Torr) exposed to ionizing radiation, was first applied in the early 1970s [8, 9], and later in the 1990s and 2000s [10-13], to measure the size distribution of ionization clusters. The basic idea was to simulate nanometer-scale ionization damage to the DNA by measuring the structure of mm-scale clusters in a tissue-equivalent gas. The ions were electrostatically extracted from the low-pressure gas volume to vacuum, where they were detected by a micro-channel or micro-sphere plate. More recently, an idea for position-sensitive detection of positive ions by impact ionization of gas molecules at low pressure (few Torr) using a Thick GEM-like structure [14] was suggested in [15]. Here the main difficulty is to accelerate the ions to sufficient energy between collisions, necessitating operation at low pressure and the use of very thick structures [16].

In what follows we discuss the possibility of positive ion detection in atmospheric and high-pressure gaseous TPCs, using potential emission of secondary electrons from surfaces by the impinging ions. The emitted electrons can be used to generate localized light or charge signals, recorded by position-sensitive readout. We note that experimental work on *internal* potential emission (where valence electrons are excited to the conduction band) by positive ions was recently reported on in [17], and that an idea of using external potential emission for positive ion detection in low-pressure TPCs for nanodosimetry studies was recently raised by V. Dangendorf independently of this proposal [18]. We begin with an overview of potential emission processes, followed by general considerations for employing them in gaseous TPCs and a discussion of several candidate materials. We then turn to look into the possible application of the idea to the detection of Xe ions in high-pressure TPCs for $0\nu\beta\beta$ searches in $^{136}$Xe, highlighting the particular requirements and challenges in this case.

## 2. Ion-induced potential emission of secondary electrons from surfaces

The neutralization of a slow positive ion on a clean solid surface occurs primarily by means of two basic mechanisms: resonance neutralization (RN) and Auger neutralization (AN) [19-24]. If the ionization energy of the ion is sufficiently large, AN can lead to the emission of a secondary electron from the surface, in addition to the one that neutralizes the ion. A secondary electron may also be emitted in the case of RN if the resonant transition is to an excited atomic state and is followed by Auger de-excitation (AD) as explained below. Both AN and RN (followed by AD) can take place on metals, semiconductors or insulators (although for the latter the picture may be more complicated [22]). Both can occur for vanishingly small ion kinetic energies - it is the potential energy recovered during neutralization which serves to eject the electrons from the surface; as such, they are referred to as *potential emission* processes, in contrast to kinetic emission processes which are driven by the ion kinetic energy (typically at the keV scale).

RN is a one-electron process which is allowed if the ground state or one of the excited states of the approaching ion is in resonance with one of the filled continuum energy levels in the solid (i.e., if they lie opposite to a filled band). RN to the ion ground state does not generally lead to the emission of a secondary electron from the surface (Figure 1A) [21]. However, when tunneling is to an excited atomic state, de-excitation of the atom can be accompanied by the simultaneous emission of a second electron from the surface – provided that its energy is larger than the vacuum level and that its velocity is directed outward (AD, Figure 1B). A necessary condition for external electron emission by AD is that the energy of the excited state relative to the ground state of the atom is larger than the work function of the solid: $E'_x > \varphi$. The maximum kinetic energy of the emitted electron, relative to the vacuum level, is: $E_k^{max} = E'_x - \varphi$.

AN is a two-electron process in which one electron tunnels out of the surface to the ion ground state, and at the same time energy and momentum are transferred to a second electron on the surface and excite it to a higher level. Unlike RN, AN can also occur if the ion ground state lies below the bottom of the filled band (Figure 1C). As in AD, this electron may then leave the surface, depending on the energy it gained and the direction of its motion. A necessary condition for external emission by AN is that the ionization energy of the ion close to the surface (i.e., the energy difference between the ground state and vacuum level) is larger than twice the work function of the solid: $E'_i > 2\varphi$. The maximum kinetic energy of the emitted electron in this case is: $E_k^{max} = E'_i - 2\varphi$ [1]. AN is possible only if the energy transferred to the surface electron can take it to an allowed state (or above the vacuum level); transitions to either filled energy levels or into the band gap (in the case of insulators) are forbidden [22]. Moreover, even for allowed transitions, external emission of the secondary electron in both AN and AD is not guaranteed, as it may remain inside the solid; this can be referred to as *internal electron emission*. Lastly, if, in addition to AN, RN to the ground state is also possible, the probability for electron emission decreases. Note that de-excitation by photon emission is negligible in AN and AD because the radiative lifetimes are larger by ~6 orders of magnitude than those of electron transitions [21].

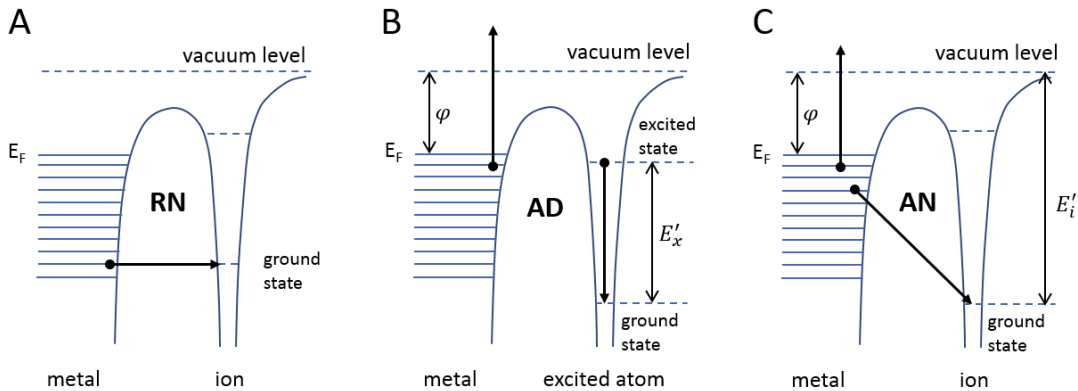

**Figure 1**: Transition diagrams for a metal target and an ion or excited (metastable) atom close to its surface. $E_F$ is the Fermi energy and $\varphi$ is the work function. (A) Resonance neutralization (RN) to the empty ground state of the ion, with no further electron emissions. (B) Auger de-excitation (AD) of an excited atom close to the surface. The de-excitation energy is simultaneously taken by a surface electron which, in this case, is excited above the vacuum level. (C) Auger neutralization (AN): an electron tunnels out of the surface to the empty ground state of the ion. The energy lost by this electron is simultaneously taken by a second electron on the surface, exciting it above the vacuum level.

---

[1] The energy levels of the ion are shifted upward close to surface (in the case of a metal) due to the interaction between the ion and its image charge; in other words, the ionization energy near the surface is somewhat lower than its free-space value. (This is the reason for writing $E'_i$ and $E'_x$ rather than $E_i$ and $E_x$.) However, this effect, which increases with the ion kinetic energy, seems to be of minor importance for slow ions [19] and may behave differently (even with an opposite sign) in insulators [22].

Ion-induced potential emission of electrons from solid surfaces was extensively studied by H. D. Hagstrum in the 1950s and 1960s, both experimentally and theoretically [19, 24-30]. He focused on singly-charged atomic ions of noble gases (He+, Ne+, Ar+, Kr+, Xe+) impinging on atomically clean metals (initially W and Mo) and on semiconductors (Si and Ge). His experiments were done in ultra-high vacuum (UHV) and the samples were atomically cleaned by flash-heating to very high temperatures (~2200 K for the refractory metals) immediately before ion bombardment. His data cover ion kinetic energies in the range 10-1000 eV. These studies, which served as the basis for all subsequent work in the field, lead Hagstrum to the advent of Ion Neutralization Spectroscopy (INS) as a sensitive probe for investigating the state densities at the solid surface by measuring the kinetic energy spectra of secondary electrons emitted by AN of slow noble gas ions [30].

Hagstrum showed that the yield $\gamma_i$ of ion-induced secondary electron emission (IISEE) (i.e., the average number of electrons emitted per ion) on atomically clean targets does not change appreciably with the ion kinetic energy over the range 10-1000 eV, supporting the interpretation that the emission is governed in this range by potential energy effects [25, 26, 28]. He demonstrated that $\gamma_i$ grows strongly with the difference $E_i - 2\varphi$ - particularly for Xe+, Kr+ and Ar+ - starting from ~1-2% for Xe+ and reaching ~30% for He+ on both Mo ($\varphi = 4.3$ eV) and W ($\varphi = 4.5$ eV). His data for Mo and W targets and 10 eV ions are shown in Figure 2, which also lists the free-space ionization energy of the ions. Later works by Hagstrum and other authors on additional atomically clean metal surfaces (Cu, Ni, Au, Pt, Ta) showed electron yields of the same order of magnitude and of similar qualitative dependence on the ionization energy, but with generally smaller values than those for Mo and W [30, 31].

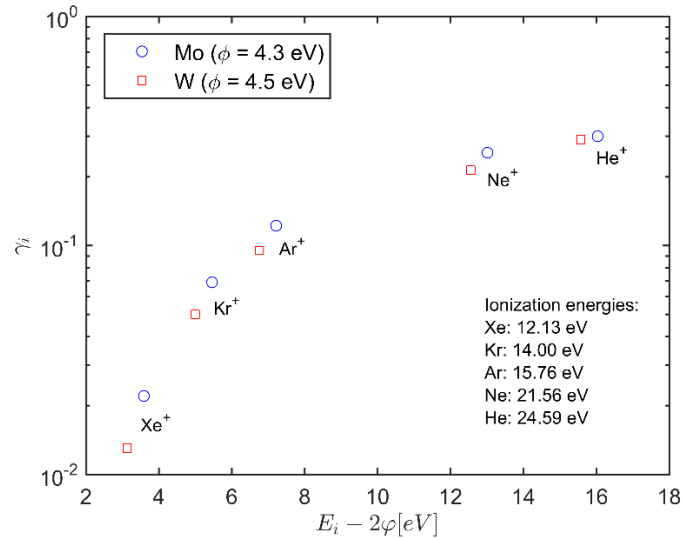

**Figure 2**: Adapted from [26] – Hagstrum's measured ion-induced secondary electron yields on atomically clean Mo and W targets, as a function of the difference between the ionization energy of the ion in free space and twice the work function of the surface. Data shown for 10 eV ions.

Hagstrum's measured IISEE yields for Ge and Si (for the same noble gas ions) were substantially smaller than those for Mo and W, by a factor which strongly increased with decreasing ionization energy (or with $E_i - 2\varphi$): ~1.5 for He+, ~2 for Ne+, ~4 for Ar+, ~10 for Kr+ and ~40 for Xe+ [28]. The work functions for the relevant crystal planes of Ge and Si are, respectively, 4.8 eV and 4.7 eV. Since these values are not much different than the work functions of Mo and W, the much lower IISEE yields were attributed to the large width of the semiconductors valence band, which results in increasing competition between AN and RN to the ground state for ions with lower ionization energy (whereas for the refractory metals AN was the only available neutralization channel) [29].

A central observation by Hagstrum was that the presence of an adsorbed foreign monolayer on the target surface substantially reduces electron emission, by a factor which, again, increases with decreasing ionization energy [27]. The effect was studied by controllably exposing an atomically clean W target to $N_2$, $H_2$ and CO during bombardment by singly-charged noble gas ions. In all cases the rate of electron emission decreased steadily during the formation of the adsorbed monolayer and leveled off at a final steady-state value. For example, for a W target with $N_2$ monolayer and 10 eV ions, the suppression factor was 1.6 for $He^+$ and $Ne^+$, 2.1 for $Ar^+$, 2.9 for $Kr^+$ and 6.5 for $Xe^+$. Electron emission could be recovered to its original value by re-heating the W sample to 2000 K in the presence of $N_2$ and CO (but not $H_2$). The suppression of IISEE by an adsorbed monolayer of foreign material was repeatedly confirmed in later works with precisely contaminated surfaces [21, 23, 31]. The effect may be partly attributed to an increase in the effective work function of the surface, as electrons are more tightly bound to the contaminating molecules, such that the condition $E_i' > 2\varphi$ is not satisfied [23].

As discussed above, Hagstrum focused on IISEE by atomic ions of noble gases. Subsequent works showed that potential emission was also possible for *molecular* ions. Vance et al [32, 33] studied this for atomically clean Mo, using $H_2^+$ ($E_i$ = 15.43 eV), $N_2^+$ (15.58 eV), $O_2^+$ (12.07 eV) and $NO^+$ (9.26 eV) with kinetic energies extending down to 30 eV. In this case, $\gamma_i$ values were considerably smaller than those for ions of noble gas atoms of similar ionization energy (3.3% for $H_2^+$, 2.3% for $N_2^+$, and 0.7% for $O_2^+$ at 30 eV, ~0.1% for $NO^+$ at 50 eV) and displayed stronger, monotonically increasing, dependence on the ion kinetic energy. This different behavior is not fully understood and may be related to the denser energy structure of the molecules compared to single atoms [20].

Compared to metals and semiconductors, IISEE experiments on insulating targets are more difficult to perform because of charging-up effects during ion bombardment. Furthermore, additional energy dissipation channels are possible, complicating the interpretation of such experiments. Nevertheless, it appears that the basic processes of RN, AD and AN also play a dominant role in the case of insulators, with electron transitions occurring from the valence band (as in semiconductors) [22]. Substantial work on IISEE from insulating films was done during the 1990s and 2000s in the context of the development of AC plasma display panels (PDP). The aim was to optimize the discharge characteristics of these devices (which were operated with various combinations noble gases, mostly Ne/Xe mixtures), by coating the inner surface of the glass-covered cathode with a thin and durable dielectric film, exhibiting a high yield of IISEE. The primary focus was on MgO, which combines high IISEE yields with a high degree of sputter resistance, and was therefore the material of choice in most PDP cells. IISEE of slow noble-gas ions (with a typical energy of a few tens of eV) on MgO was measured by several groups, showing a large spread in $\gamma_i$, as summarized in [34]: $He^+$ 13-35%, $Ne^+$ 6-50%, $Ar^+$ 0.9-6%, $Kr^+$ 0.9-1.5%, $Xe^+$ 0.2-4%. This spread is attributed to complications in the experimental methodologies employed, the quality of the MgO surface and the studied crystal plane (with MgO(111) giving the highest yields) [22]. Experiments were done either using a beam of slow ions in UHV setups (as was done by Hagstrum on metals and semiconductors), or in discharge experiments in which $\gamma_i$ was inferred from measuring the breakdown voltage in noble gases at sub-atmospheric pressures. The former line of experiments is complicated by charging-up effects, and the latter by the difficulty in separating the contribution to secondary electron emission from ions (of different species), photons and metastable atoms. A particularly careful study was carried out by Matulevich et al [35,36] on 1-5 nm thick MgO on Mo(001) using a beam of 40 eV singly-charged noble gas ions with very low current to avoid charging-up effects. The measured yields in this case were: $He^+$ 55 ± 5%, $Ne^+$ 41 ± 4%, $Ar^+$ 10 ± 1% and $Xe^+$ 3 ± 2%. Based on theoretical model simulations the authors concluded that these values are too high to be explained by AN alone; this is particularly evident for Xe, where the condition $E_i' > 2\varphi = 2(E_g + \chi)$ is not met (at the MgO surface $E_g$ = 6.8 eV and $\chi$ = 0.85 eV [34], where $E_g$ is the width of the band gap and $\chi$ is the electron affinity). They suggested that electron emission is enhanced by a process in which the remaining hole is transported across the thin MgO layer towards the Mo substrate, where its neutralization by an Auger transition results in the emission of another electron that may leave the surface.

Several groups investigated the suitability of other dielectric layers as secondary electron emitters in the context of PDP development. In particular Bachmann *et al* [37, 38] inferred secondary electron yields for hydrogen-terminated CVD diamond from discharge experiments with Ne, Ar and Xe, finding values of ~50%, ~20% and ~2%, respectively. The authors discuss the possibility that other processes (such as electron emission by UV photons) contribute to the measured yield, but conclude that it is dominated by ion-induced emission. The study compared between hydrogen-terminated CVD diamond (with negative electron affinity NEA) and oxygen-terminated CVD diamond (with positive electron affinity), showing much higher yields for the former. The NEA value reported is -0.8 eV [37], giving (for a band gap of 5.5 eV) a work function of 4.7 eV.

Lastly, very high $\gamma_i$ values were inferred from measured gain curves by Lyashenko *et al* [39] for bi- and mono-alkali photocathodes (K–Cs–Sb, Na–K–Sb, and Cs–Sb) in 700 Torr Ar/5%CH$_4$. In this case *thermal* ions (presumably $CH_4^+$) impinging on the photocathodes (with extremely low work functions ranging from 0.5 to 1.1 eV) resulted in $\gamma_i$ values between 47% and 49% (with large uncertainties).

## 3. General considerations for employing ion-induced potential emission in TPCs

Since potential emission can occur for thermal ions (as demonstrated in [39]), it can conceivably be used to detect positive ions in gaseous TPCs operated at atmospheric or high pressure, thus opening new experimental possibilities. The basic process would be the release of secondary electrons by ions reaching the cathode plane, with subsequent detection of the emitted electrons through charge or light signals. For this to become a reality, however, several important considerations should be taken into account.

*Surface purity:* As discussed above, the presence of an adsorbed monolayer of foreign molecules on the surface of the target can drastically reduce the IISEE yield. It therefore appears that IISEE target materials could only be used in ultra-pure gases which either do not stick to the target surface or can be easily removed. Even then, rapid contamination of the surface seems unavoidable. Consider, for example, a TPC operated with a pure noble gas at 1 bar, containing impurities at the ppb level. The partial pressure of these impurities will be ~$10^{-6}$ Torr which, in vacuum, would result in the formation of a monolayer within several seconds. It seems that in order to keep the surface clean, periodic heating to high temperatures will be mandatory. This could be done, for example, by depositing the target material on thin resistive wires or strips (on an insulating substrate) and apply flash ohmic heating. The wires themselves, of course, could serve as the target material (e.g., Mo or W wires). In addition to periodic flash-heating, the wires could be heated on demand following a trigger (for example a prompt scintillation-based $t_0$, and/or electron-based signal coming from the anode), making use of the slow drift velocity of the ions to have the target surface clean at the expected ion time of arrival. The required temperatures and periodicity of the heating cycles will vary between different types of IISEE target materials, and may expected to be lower for chemically-inert surfaces, such as MgO or H-terminated diamond. When designing such heating cycles, care should be taken to minimize convective flows inside the TPC that may interfere with the ion drift.

*Ion species reaching the cathode:* If the TPC contains a mixture of several gases, due to collisional charge exchange the positive ions reaching the cathode will be those of the species that has the lowest ionization energy. In pure noble-gas TPCs ions formed along the ionization track quickly convert through three-body collisions to molecular (dimer) ions through the reaction: $X^+ + 2X \rightarrow X_2^+ + X$ [40], which have slightly lower ionization energy compared to the atomic ions [34]. Although, to the best of our knowledge, no experimental data is available on the IISEE yields for noble gas dimer ions, they are likely lower than those of the corresponding atomic ions (as was observed for other molecular ions). The picture may be further complicated by the formation of clusters of neutral atoms around the drifting ions [41-45]. This can lead to dispersion in drift velocities if there is more than one dominant cluster type. The importance of this effect may vary greatly between different gases and operating pressures and should be studied in detail for any particular application.

*Surface work function:* As discussed above, the IISEE yield strongly increases with the difference between the ionization energy and twice the work function. For most singly-charged ions (except He$^+$ and Ne$^+$) target materials with low work function, preferably ~3-4 eV or below, will be necessary for having an appreciable detection efficiency.

*Ion detection efficiency and potential applicability*: The maximum reported values of $\gamma_i$ are ~50%. For many ion/target combinations yields on the few-percent level should be expected. Backscattering of the emitted electrons by elastic collisions within the gas will further reduce the effective yield in most cases. This apparently limits the potential scope of applicability to experiments in which track reconstruction can be carried out using a fraction of the primary charge, or where useful information can be obtained from the ions time of arrival. Note that the information extracted from the track's positive ions should, in most circumstances, come in addition to that gathered from its electrons.

*Electron extraction into gas:* As noted above backscattering by elastic collisions with the gas molecules will reduce the effective yield, which can be expressed as $\gamma_{eff} = \gamma_i \cdot \varepsilon_{ext}(gas; E/p, \bar{E}_k)$ [39]; $\varepsilon_{ext}$ is the extraction efficiency into gas, i.e., the probability that the emitted electron is not backscattered to the cathode. The extraction efficiency depends on the gas, the reduced field $E/p$ on the cathode surface and the average kinetic energy of the emitted electrons, $\bar{E}_k$. At a given reduced field, it is generally low for pure noble gases, for which elastic collisions are dominant at low kinetic energies, and higher for molecular gases where inelastic scattering is important. The extraction efficiency increases monotonically with $E/p$ but tends to saturate at a value which may be considerably lower than 1. For some gases (e.g., Ar, Kr, Xe) it *decreases* with the kinetic energy of the emitted electron because the elastic scattering cross section increases with energy (for kinetic energies above the Ramsaeur–Townsend minimum of the elastic cross-section) [46]. In this case a lower work function will not necessarily lead to a higher effective IISEE yield because the kinetic energies of the emitted electrons will be higher, resulting in lower extraction efficiency.

*Electron field emission:* Low work function materials may also show a high probability for non-thermal electron field emission (EFE). Since ions arrive one by one and emit at most one electron per neutralization, individual EFE events will look the same as IISEE events.

*Readout scheme:* The readout system employed must be able to detect individual electrons emitted by the impinging ions. If the detection is based on avalanche multiplication, the required charge gain should be high (~$10^4$-$10^5$) depending on the readout electronic noise. This will be difficult to achieve in TPCs with a pure noble gas, requiring the addition of a quencher to avoid discharges. As discussed below for the case of high-pressure Xe, single-electron sensitivity appears to be possible by light detection, in a structure combining small charge gain and electroluminescence (EL). This can be generalized to other gases with high EL yields.

## 4. Candidate IISEE target materials for gaseous TPCs

The ideal IISEE material should combine several properties: (1) low work function (~3-4 eV or below); (2) low probability of electron field emission (EFE); (3) high degree of chemical stability, or (4) suitability for *in situ* surface cleaning by flash heating; (5) suitability for thin wire/strip geometries.

Mo and W appear to be the most suitable candidate metallic IISEE materials. For both metals (when their surface is atomically clean) AN is considered to be the only available neutralization channel, with no competition from RN to the ion ground state; this results in IISEE yields ranging from 2% for Xe$^+$ on Mo, through ~10% for Ar$^+$ to ~30% for Ne$^+$ and He$^+$ for both metals - in spite of their somewhat high work function (~4.3-4.5 eV for the proper crystal planes). Both Mo and W (especially W) are suitable for forming wire planes, and both are suitable for *in situ* flash ohmic heating to very high temperatures – which is likely unavoidable as their surface is not chemically inert. For ions of low ionization energy Mo appears to be preferable over W. There is no known EFE problem with either Mo or W.

MgO may be a promising option – especially if applied as an ultra-thin layer on Mo. As noted above, 1-5 nm thick MgO on Mo displayed very high IISEE yields: 55% for He$^+$, 41% for Ne$^+$, 10% for Ar$^+$ and 3% for Xe$^+$, in spite of the fact that for Xe the AN condition $E_i' > 2\varphi$ is not satisfied (assuming the nominal work function of MgO). While the investigators suggest that this may reflect an enhancement of AN by an additional mechanism [35, 36], it may also originate from an effective reduction in the Mo work function. A recent study demonstrated a work function of 3.2 eV for an 8-atomic layer thick film of MgO on Mo(001) [47], while density functional calculations suggest that applying 1-3 atomic layers of MgO on Mo(001) can lead to a work function of 2.05 eV [48]. In addition to the possibility of a very low work function, MgO surfaces are essentially chemically inert – making it an apparently ideal IISEE candidate; if some *in situ* heating is required, it can be done easily by ohmic heating of the Mo substrate. It remains to be demonstrated that ultra-thin MgO films on Mo wires or thin strips can have similar properties with no EFE issues. As long as the use of MgO is limited to low-rate (e.g., rare-event) experiments, charging up effects are expected to be minimal.

Diamond layers with negative electron affinity (NEA) are an additional interesting option. As discussed above, high IISEE yields were inferred from discharge studies for H-terminated CVD diamond with NEA of -0.8 eV (i.e., work function of 4.7 eV), namely ~50% for Ne, ~20% for Ar and ~2% for Xe [37, 38]. Since in this case RN to the ion ground state is possible (and therefore competes with AN without electron emission), these high values may have been affected, to some extent, by additional mechanisms such as photoemission, which can be quite large for the relevant VUV wavelengths [49]. A recent study [50] demonstrated a very large NEA of -2.0 eV for Mg adsorbed on oxygen-terminated single-crystal diamond (100), with a work function changing from 2.6 to 3.8 eV after successive steps of annealing at increasingly higher temperatures; exposure to air increased the work function to 4.4 eV (going down to 4.2 eV after heating to 250°C) and immersion in water resulted in a minor increase of the work function to 4.5 eV, displaying a very high degree of surface inertness. New techniques in growing nanocrystalline diamond (NCD) and ultrananocrystalline diamond (UNCD) [51], provide a high degree of flexibility in tuning the properties of the layer for a wide range of applications. In particular, hydrogen termination of nitrogen-incorporated 150 nm thick layer of UNCD ((N)UNCD:H) was shown to provide a stable work function of 3.1 eV, as well as protection against exposure to air [52]. A potential problem with UNCD, however, is a high degree of EFE [53-55], which may be prohibitive for large surface fields (tens of kV/cm); it remains to be demonstrated if a sufficiently strong EFE suppression is possible by tuning the UNCD parameters without, at the same time, paying a penalty of a higher work function. EFE is reported to be very low in single-crystal diamond [55], but it is unclear if this is a practical option for TPCs. The surface treatment of the above diamond layers provides a sufficiently high conductivity to avoid charging-up.

Mono- and bi-alkali photocathodes are perhaps the most sensitive IISEE materials, due to their exceedingly small work functions. At the same time, they are extremely sensitive to exposure to low levels of humidity and cannot be heated. Their use in rare-event TPCs, which must typically take data for several years, therefore seems highly challenging. However, one can imagine using such photocathodes in ultra-pure setups for short-term dedicated experiments aimed at detecting ions in well-defined processes (e.g., mobility measurements of ionic species with low ionization energy).

## 5. The case of $0\nu\beta\beta$ in $^{136}$Xe

The concept of a high-pressure Xe (HPXe) TPC for $0\nu\beta\beta$ searches in $^{136}$Xe, developed by the NEXT collaboration [56] and recently adopted by PandaX-III [57], has two unique features: excellent energy resolution (0.5-0.7% FWHM at $Q_{\beta\beta} = 2.46$ MeV), and track reconstruction capability. High energy resolution is critical for rejecting $2\nu\beta\beta$ events with energy close to $Q_{\beta\beta}$, while accurate imaging can provide effective discrimination against gamma-induced background events, based on the track topology. In the absence of a magnetic field, the main difference between a candidate ~20 cm long $0\nu\beta\beta$ track and a single-electron track is that the former has two 'blobs' of dense ionization at its opposite ends, while the latter generally has only one. The ability to extract topological information from ionization tracks was demonstrated in [58] using the NEXT-DEMO prototype, and the possibility of employing deep neural network algorithms to further enhance topology-based background rejection was demonstrated (on simulated tracks) in [59].

Current limits on $0\nu\beta\beta$ halflives indicate that the mass of future experiments should be on the ton-scale. The drift length $L$ of a TPC containing 1 ton of Xe at 10 bar will be ~3 m (for equal diameter and length) or ~1.5 m in a bidirectional design. For a realistic field $E = 300$ V/cm the r.m.s. electron diffusion in pure Xe over 1.5 m drift is $\sigma_{el} \approx 16$ mm [2]. This will likely wash out the track blob structures, thus seriously affecting background rejection. Electron diffusion can be reduced by small admixtures of polyatomic gas, but this results in lower energy resolution. For example, simulations indicate that adding 0.5% $CH_4$ to pure Xe will reduce $\sigma_{el}$ over 1.5 m from 16 to ~3 mm, but at the price of 1.6 larger FWHM [2].

Detection of the positive ions of a candidate $0\nu\beta\beta$ track *in addition* to its electrons could potentially enable accurate track imaging along with the excellent energy resolution of pure Xe. In this scheme (Figure 3) the TPC cathode plane is modified to detect positive ions, while the anode plane still includes an EL structure (e.g., two parallel meshes, or a THGEM-like structure). Both the cathode and anode planes are equipped with position-sensitive arrays of photon detectors. Primary scintillation is recorded on both planes, providing a clear $t_0$ signal. The electrons reach the anode plane within several hundred μs to produce an EL signal for precise measurement of the energy. A gross image of the track is recorded by the photon detectors at the anode plane. The ions reach the cathode several seconds later, where their position and time of arrival are recorded with high accuracy; the gross electron-based image recorded at the anode serves to narrow the data acquisition window for ion detection in terms of time and position.

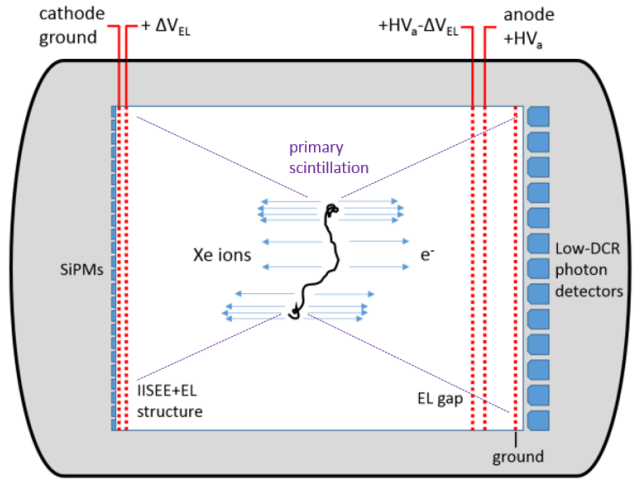

**Figure 3**: Proposed (one-directional) scheme for HPXe TPC for $0\nu\beta\beta$ searches in $^{136}$Xe. Positive Xe ions are detected by ion-induced secondary electron emission (IISEE) followed by electroluminescence (EL) on the cathode plane, with accurate imaging by a dense array of SiPMs. Electrons produce an EL signal at the anode for energy measurement and gross imaging by the nearby photon detectors. Primary scintillation for $t_0$ determination is recorded on both planes (but more easily extracted from the low dark-count-rate detectors on the anode side). A bidirectional scheme where the design is mirrored with respect to a central cathode plane is also possible.

The cathode plane comprises two elements: an ion-induced EL structure and a dense array of pixelated detectors (e.g., silicon photomultipliers, SiPMs) closely behind. Two possible options are depicted in Figure 4: (A) a micro-strip gas counter (MSGC) pattern on tiled modules of fused silica plates, made up of ion-collecting IISEE strips (e.g. MgO or NEA diamond on Mo) interlaced with electron-collecting metal strips; (B) two wire planes, where the one closer to the photon detectors (i.e., the cathode) includes Mo or W wires coated with MgO or NEA diamond, and the other serves to define the field. In (A), ions landing on the IISEE strips lead, with some probability, to the emission of secondary electrons; these are extracted into gas by the strong surface field, undergo modest avalanche multiplication, produce a localized pulse of EL and land on the neighboring metal strips; the same occurs in (B), where the emitted electrons are collected on the second wire plane. In both configurations the IISEE strips and wires can be ohmically flash-heated following a trigger based on $t_0$ and the anode EL signals; this will allow for precise timing of the flash-heating current pulse to bring the strips to a desired temperature when the ions arrive. To facilitate flash-heating and minimize the distance between the IISEE+EL structure and photon-detection plane, the cathode will be at ground potential, with the anode at +HV. This will require a ~10 cm gap between the anode EL structure and its photon-detector array (which should be further screened by a grounded mesh).

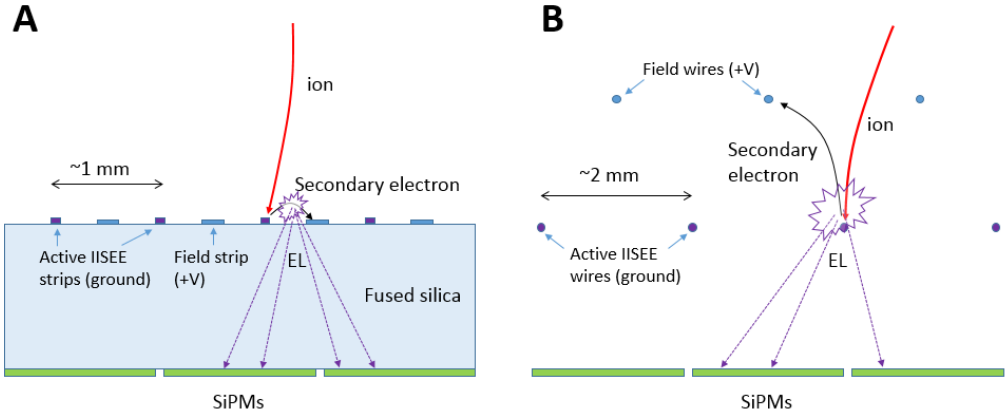

**Figure 4**: Possible ion detection structures on the cathode side (here drawn horizontally). (A) MSGC-like structure, with ion-collecting active IISEE strips interlaced with electron-collecting metal strips, deposited on a fused silica plate. Secondary electrons emitted by arriving ions from the active strips undergo modest avalanche multiplication and produce EL light - detected by a dense array of SiPMs – and are collected by the metal field strips. (B) The same principle, implemented with parallel planes of active IISEE wires and field wires. Dimensions are not to scale.

The photon detectors of the anode plane should probably not be SiPMs, to simplify the detection of the $t_0$ signal above the dark count noise; instead, one can consider a closely packed array of photomultiplier tubes (PMTs) or an array of large, highly-pixelated, gaseous photomultipliers (GPMs) behind fused silica windows [60,61]. Both options will provide a very low dark count rate (DCR). For GPMs fine granularity will likely be required for maintaining high energy resolution for the anode EL signal, by photon-counting on individual pixels; this is needed because of the exponential single-electron pulse height distribution of gas detectors.

We now turn to semi-quantitative estimates regarding different aspects of the proposed scheme. A candidate $0\nu\beta\beta$ event produces $\sim 1 \times 10^5$ electron-ion pairs which escape recombination [7]. The $Xe^+$ ions immediately (~0.1 ns) convert in three-body collisions to $Xe_2^+$ dimer ions [40], which may quickly convert to $Xe_3^+$ or larger clusters [41-44]. Xe neutral atom clustering around Xe ions is not sufficiently studied at 10 bar [7]; hopefully clustering results in one dominant ion species. If several cluster sizes coexist there will be a spread of ion drift velocities, smearing out the track along the drift direction; this, however, should not directly affect the reconstructed *xy* projection of the track. Extrapolating from measured values of the reduced mobility of $Xe_3^+$ at 0.07 bar [41], the expected drift velocity at 10 bar under 300 V/cm would be ~10-15 cm/s. Careful measurements of Xe cluster mobility at high pressure are obviously of vital importance here.

The ionization energy of $Xe_2^+$ is 11.18 eV [62]. Data for larger clusters is less clear, but it appears that the ionization energy is roughly constant regardless of the cluster size $n$ for $n > 2$, with values around 10.5-11 eV [63, 64]. Since the gas is continuously purified, and since all common impurities ($H_2$, $H_2O$, $O_2$, $N_2$, $CO_2$, CO, $CH_4$) have ionization energies larger than 12 eV, the Xe ion clusters can be expected to reach the cathode without being neutralized by charge-exchange collisions along their drift.

From Einstein's relation, the r.m.s. diffusion of thermal ions (irrespective of their mass) is given by:

$$\sigma_{ions} = \sqrt{2Dt} = \sqrt{\frac{2k_B T}{q_e} \cdot \frac{L}{E}} = 2.25\sqrt{\frac{L}{E}} \qquad (1)$$

where $\sigma_{ions}$ is in mm, $L$ in cm and $E$ in V/cm. For $L = 150$ cm, $E = 300$ V/cm this gives $\sigma_{ions} = 1.6$ mm, a 10-fold improvement compared to electrons (note that this is the *maximal* spread, for ions crossing the entire TPC). The spacing between the IISEE strips or wires should preferably be smaller than this to keep the track image as sharp as possible. For wire planes, the realistic distance between wires is probably limited to ~2 mm. For strips on fused silica the spacing can be smaller; however, going below ~1 mm between active strips (0.5 mm between active and electron-collecting field strips), may limit the EL yield unless substantial charge gain is allowed. Preliminary calculations which assume an overall photon detection efficiency of 10% for the cathode SiPM array (with 6 mm square pixels 3 mm from the strips) show that in order to reconstruct the ion arrival point with ~0.5 mm r.m.s. uncertainty, the EL pulse it induces should comprise ~2000 photons emitted into $4\pi$ (resulting in ~100 detected photons). Given the small distances between strips/wires this should be possible if the surface field is large enough to allow an amplification charge gain of ~5-10 (requiring ~200-400 photons per electron). Gain fluctuations seem unimportant as the amplification is not aimed at energy measurement. Careful optimization is required, but the numbers appear feasible with some flexibility in the geometry and applied voltages. The DCR of current generations of SiPMs at room temperature is $\sim 5 \times 10^4$ Hz/mm$^2$. If the EL signal lasts $\lesssim 100$ ns (where most of the light is emitted ~1 mm from the strip/wire), and 100 photons are detected over an active SiPM area of ~500 mm$^2$, the signal-to-noise ratio will be ~40. Care should be taken not to increase the SiPM temperature during flash-heating of the strips/wires. Alternatively, if DCR becomes an issue, one may consider modest cooling of the SiPM plane.

The main unknown in the proposed scheme is the IISEE yield for Xe ion clusters impinging on the active material. Since for molecular ions the yield is smaller than for atomic ions of similar ionization energy [32, 33], the expected yield for materials with a work function of ~4.3-4.7 eV (Mo, W, diamond with modest NEA) will likely be considerably smaller than 2%. This, however, may significantly improve for materials with a work function of ~3 eV (e.g., few atomic layers of MgO on Mo, or (N)UNCD:H). Since the strips/wires will be flash-heated upon trigger (and can remain at elevated temperature when the ions arrive), the presence of an adsorbed monolayer of either molecular impurities or physisorbed Xe atoms can be avoided. As noted above, care should be taken not to induce large convective flows during heating.

The probability of IISEE into the gas is the effective yield: $\gamma_{eff} = \gamma_i \cdot \varepsilon_{ext}(Xe; E/p, \bar{E}_k)$. Monte Carlo calculations on photoelectron emission from CsI into pure Xe, under irradiation at 185 nm (photon energy of 6.7 eV, corresponding to a maximum initial electron kinetic energy of ~0.6 eV [65]), show that $\varepsilon_{ext} \approx 0.2$ already at $E/p = 1$ Vcm$^{-1}$Torr$^{-1}$ (corresponding to 7.5 kV/cm at 10 bar) [66]. (The simulations were validated experimentally on Xe/CH$_4$ mixtures down to 1% CH$_4$.) Preliminary Garfield++ [67] Monte Carlo calculations of the extraction efficiency into Xe at 10 bar are consistent with [66] and indicate that it does not increase significantly at higher fields; these calculations further show that for a given field $\varepsilon_{ext}$ decreases with the initial kinetic energy of the emitted electron.

In order to accurately reconstruct the topology of a candidate $0\nu\beta\beta$ event it appears that several hundred detected ions are sufficient. For example, if 200 ions are detected, the typical spacing between adjacent ions for a 20 cm long track will be ~1 mm, with higher densities at the opposite ends of the track. For $1 \times 10^5$ ions arriving at the cathode, this will require $\gamma_{eff} = 0.2\%$ (e.g. by having, $\gamma_i = 1\%$ and $\varepsilon_{ext} = 0.2$). Higher values for $\gamma_{eff}$ will obviously improve the quality of the image (which will inevitably contain some noise). Note that in addition to identifying the blob structures, further information on the track curvature (which should be readily accessible with $\gtrsim 200$ detected ions) may assist in background discrimination [68].

The main source of noise will be random electron field emissions from the IISEE strips/wires. Assuming 1 detected ion per mm$^3$ of track, and a drift velocity of ~10 cm/s, an overall EFE rate of 10 Hz/mm$^2$ (over the cathode surface) will result in a signal-to-noise ratio of 10. If the cathode plane is made up of 20 μm-wide strips with a spacing of 1 mm, this corresponds to an EFE rate of 500 Hz per mm$^2$ of active IISEE material (into Xe at 10 bar). Assuming $\varepsilon_{ext} = 0.2$ this translates in vacuum to EFE current density of $4 \times 10^{-14}$ A/cm$^2$ of active IISEE material. This appears to be a challenging requirement of UNCD [53-55], but may possibly be overcome by tuning the layer's parameters. As noted above, there are no known data on EFE from ultra-thin MgO on Mo.

Photon feedback from the IISEE strips/wires may potentially become an issue if the quantum efficiency of the active material at 172 nm is large. However, the risk of uncontrolled feedback can be mitigated by optimizing the geometry and voltages of the IISEE-EL structure – e.g., by reducing the ratio between the strip width and spacing, by reducing the surface field on the active strips and by allowing part of the avalanche to develop next to the collecting strips (with similar consideration for wire planes).

## 6. Concluding remarks

Potential emission of secondary electrons by positive ions may open new experimental avenues in low-rate gaseous TPCs. Since the basic mechanism can work for thermal ions, it may be applied in TPCs operating at atmospheric or high pressure. The intrinsic spatial resolution is governed by thermal diffusion and can be on the sub-mm scale, depending on the TPC size and field strength. In principle, the proposed method can allow the detection of individual ions, one-by-one. The key requirements from the active materials are a low work function, high degree of chemical inertness and/or suitability for *in situ* heating, and a low rate of electron field emission.

Considering the case of $0\nu\beta\beta$ searches in $^{136}$Xe using a HPXe TPC, positive ion detection may allow ~1-2 mm accuracy in track reconstruction without compromising the excellent energy resolution of pure Xe. It appears that the main challenge will be identifying an electron emitting material that offers an IISEE yield of a few percent for Xe ion clusters, accompanied by very low electron field emission. Luckily, there appear to be viable materials that can serve as a starting point for this search. A potential complication is the possibility that several sizes of Xe ion clusters coexist with similar concentrations at high pressure. This may interfere with track reconstruction along the drift direction, but should not directly affect the accuracy in *xy*. Other R&D issues include the (likely) necessity to flash-heat the IISEE strips/wires following a trigger from $t_0$ and the anode EL signal, and the multivariate design of the IISEE+EL structure and readout. Finally, realistic estimates should be made regarding the impact of the proposed scheme on the overall sensitivity of the experiment for $0\nu\beta\beta$ detection.

Conceivably, as noted by D. Nygren in [7] with respect to $0\nu\beta\beta$, and by N. Spooner with respect to directional dark matter searches [69], one may consider incorporating positive ion detection into negative ion TPCs. Considering the efficiency of positive ion detection, the potential merit of this will be the addition of an 'echo' signal originating from electron release from the cathode when the ions arrive, enabling to position the event in three dimensions.

Experimental measurements of IISEE yields and field emission rates of candidate materials, with particular emphasis on Xe, are now underway. As a directly relevant byproduct, these experiments are expected to provide valuable information on ion cluster mobilities at different pressures.

The challenges associated with harnessing ion-induced potential emission to gaseous TPCs are varied and numerous, and addressing them will likely unravel new ones. Yet, in view of the rapid advances in material and surface science, one can hope that this dream may come true.

**Acknowledgements**

I would like to thank A. Breskin for his ongoing support and encouragement, as well as for introducing me to past works of relevance to this proposal; D. Nygren for the stimulating discussions that eventually led to this idea, and for many insightful comments along the way; S. Shchemelinin for eye-opening discussions on potential emission; A. Para for the enthusiastic Ping-Pong of comments and ideas and for initiating the connection with ANL, and A. Sumant and A. Hoffman for the discussions on diamond layers and for their willingness to join the experimental effort. This work is supported by the Weizmann Institute Staff Scientist Internal Grant Program.